\documentclass[prc,aps,twocolumn,amssymb,aps,nofootinbib,floatfix]{revtex4}
\usepackage[ansinew]{inputenc}
\usepackage[T1]{fontenc} 
\usepackage{graphicx}
\usepackage{epsfig}

\begin{document}

\title{Strangeness fluctuations and MEMO production at FAIR}

\author{Jan Steinheimer${}^1$, Michael Mitrovski${}^{1,2}$, Tim Schuster${}^{1,2}$, Hannah Petersen${}^{1,2}$, Marcus Bleicher${}^1$ and Horst St\"ocker${}^{1,2,3}$}
\affiliation{${}^1$ Institut f\"ur Theoretische Physik, Johann Wolfgang Goethe-Universit\"at, 
Max-von-Laue-Str.~1, 60438 Frankfurt am Main, Germany}
\affiliation{${}^2$ Frankfurt Institute for Advanced Studies (FIAS), Johann Wolfgang Goethe-Universit\"at, 
Ruth-Moufang-Str.~1, 60438 Frankfurt am Main, Germany}
\affiliation{${}^3$ GSI - Helmholtzzentrum f\"ur Schwerionenforschung mbH, Planckstr. 1, Darmstadt}

\begin{abstract}
We apply a coupled transport-hydrodynamics model to discuss the production of multi-strange meta-stable objects in Pb+Pb reactions at the 
FAIR facility. In addition to making predictions for yields of these particles we are able to calculate particle dependent rapidity and momentum distributions. We argue that the FAIR energy regime is the optimal place to search for multi-strange baryonic object (due to the high baryon density, favoring a distillation of strangeness). Additionally, we show results for strangeness and baryon density fluctuations. Using the UrQMD model we calculate the strangeness separation in phase space which might lead to an enhanced production of MEMOs compared to models that assume global thermalization.
\end{abstract}

\maketitle
Massive heavy-ion reactions provide an abundant source of strangeness. More than 50 hyperons 
and about 30 Anti-Kaons (i.e. $K^-+\overline{K^0}$ carrying the strange quark)
are produced in central collisions of lead nuclei at the CERN-SPS low energy program and before that at
the AGS (see e.g.\ \cite{SQM98}). In the near future, the Facility for Anti-proton and Ion Research (FAIR)
will start to investigate this energy regime closer with much higher luminosity and
state-of-the-art detector technology. This opens the exciting perspective to explore the formation 
of composite objects with multiple units of strangeness so far unachievable with conventional methods.

Exotic forms of deeply bound objects with strangeness have been proposed long ago (see \cite{Bodmer:1971we}) 
as collapsed states of matter, either consisting of baryons or quarks.  
For example the H di-baryon (a six quark state) was predicted by Jaffe
\cite{Jaffe:1976yi}. Later a multitude of bound di-baryon states with strangeness were proposed
using quark potentials \cite{Goldman:1987ma,Goldman:1998jd} or the Skyrme model \cite{Schwesinger:1994vd}.
However, the (non-)observation of multi-quark bags, e.g. strangelets 
and (strange) di-baryons is still one of the open problems of 
intermediate and  high energy physics. Most noteworthy in this respect has been the hunt for the Pentaquark over the last 10 years, which re-stimulated this field and resulted in a reported observation at the CERN SPS accelerator \cite{Alt:2003vb}.\\
The early theoretical models based on SU(3) and SU(6) symmetries \cite{Oakes:1963zz,Dyson:1964} and on Regge theory
\cite{Libby:1969tw,Graffi:1969is} suggest that di-baryons should exist. More recently,
even QCD-inspired models predict di-baryons with strangeness S = 0, -1, and -2. The
invariant masses range between 2000 and 3000 MeV \cite{Jaffe:1976yi,Aerts:1977rw,Wong:1978tz,Aerts:1984ht,Kalashnikova:1987rma,Goldman:1989zj,SchaffnerBielich:1999sy,SchaffnerBielich:2000nd}.
Unfortunately, masses and widths of the expected 6-quark states differ considerably for these
models. Nevertheless, most  QCD-inspired models predict di-baryons and none
seems to forbid them.\\

On the conventional hadronic side, however, hypernuclei are known to exist already for a long time \cite{Ahn:2001sx,Takahashi:2001nm}. 
The double $\Lambda$ hypernuclear events reported so far are closely related to
the H di-baryon \cite{Dalitz:1989kt}. Metastable exotic multi-hypernuclear objects (MEMOs) 
as well as purely hyperonic systems of $\Lambda$'s and $\Xi$'s
were introduced in \cite{Schaffner:1992sn,Schaffner:1993nn} as the hadronic counterparts to
multi-strange quark bags (strangelets) \cite{Gilson:1993zs,SchaffnerBielich:1996eh}.
Most recently, the Nijmegen soft-core potential was extended to the 
full baryon octet and bound states of $\Sigma\Sigma$, $\Sigma\Xi$, and $\Xi\Xi$ 
di-baryons were predicted \cite{Stoks:1999bz}. 
For previous estimates of strangelet production and MEMO formation, the reader is referred to \cite{Scherer:2008zz,Aichelin:2008mi}.

A major uncertainty for the detection of such speculative states is their
(meta)stability. Metastable exotic multi-hypernuclear objects (MEMOs), for example,
consist of nucleons, \( \Lambda  \)'s, and \( \Xi  \) and are stabilised
due to Pauli's principle, blocking the decay of the hyperons into nucleons. 
Only few investigations about the weak
decay of di-baryons exist so far (see \cite{SchaffnerBielich:2000nd} for a full
discussion and new estimates for the weak nonleptonic decays of strange di-baryons):
In \cite{Donoghue:1986zd}, the $H$-di-baryon was found to decay dominantly 
by \(H \rightarrow \Sigma ^{-}+p \) for moderate binding energies. While the \( (\Lambda \Lambda ) \) bound state, which has
exactly the same quantum numbers as the $H$-di-baryon, was studied in \cite{Krivoruchenko:1982xv}.
Here, the main non-mesonic channel was found to be \( (\Lambda \Lambda ) \rightarrow \Lambda +n \).
If the life time of the \( (\Lambda \Lambda ) \) correlation or $H^0$ particle is not too long, 
the specific decay channels might be used to distinguish between both states.

There are several searches in heavy-ion collisions for the $H$-di-baryon \cite{Belz:1995nq,Crawford:1998uq,Caines:1999ia}
and for long-lived strangelets \cite{Appelquist:1996qx,Armstrong:1997sc} with high sensitivities, so far with no conclusive results. In $pN$ collisions at the Fermilab however, the $H$-di-baryon seems to be excluded over a wide range of Masses ($2.194 < M_H < 2.231$ GeV) and lifetimes ($5 \cdot 10^{-10}$ to $1 \cdot 10^{-3}$ sec) \cite{AlaviHarati:1999ds}.
Hypernuclei have been detected most recently in heavy-ion reactions at the AGS
by the E864 collaboration \cite{Finch:1999ja}. 
\section{MEMO production rates}
In this paper we study the production rate of multi-strange objects 
within the UrQMD model (v2.3) and a micro+macro hybrid approach to heavy 
ion reactions.

Similar to the RQMD model \cite{Stoecker:1994ud,Sorge:1989dy} which was employed in \cite{SchaffnerBielich:2000nd}, UrQMD is a microscopic transport approach based on the covariant propagation of constituent quarks and 
diquarks accompanied by mesonic and baryonic degrees of freedom.
It simulates multiple interactions of ingoing and newly produced 
particles, the excitation and fragmentation of color strings and the 
formation and decay of hadronic resonances. 
At RHIC energies, the treatment of sub-hadronic degrees of freedom is
of major importance. In the UrQMD model, these degrees of freedom enter via
the introduction of a formation time for hadrons produced in the 
fragmentation of strings \cite{NilssonAlmqvist:1986rx,Andersson:1986gw,Sjostrand:1993yb}.
The leading hadrons of the fragmenting strings contain the valence-quarks 
of the original excited hadron. In UrQMD they are allowed to
interact even during their formation time, with a reduced cross section
defined by the additive quark model, 
thus accounting for the original valence quarks contained in that hadron \cite{Bass:1998ca,Bleicher:1999xi}.

For the microscopic+macroscopic calculation, the Ultra-relativistic Quantum Molecular Dynamics Model 
(UrQMD) is used to calculate the initial state of a heavy ion collision for a subsequent hydrodynamical
 evolution \cite{Steinheimer:2007iy,Bass:1998ca,Bleicher:1999xi}. This has been done to 
account for the non-equilibrium dynamics in the very early stage of the collision. In this 
configuration the effects of event-by-event fluctuations of the initial state are naturally 
included. The coupling between the UrQMD initial state and the hydrodynamical evolution 
proceeds when the two Lorentz-contracted nuclei have passed through each other. 
\begin{equation}
t_{start} = \frac{2R}{\sqrt{\gamma^2 -1}}	
\end{equation}

After the UrQMD initial stage, a full (3+1) dimensional ideal hydrodynamic evolution is performed 
using the SHASTA algorithm \cite{Rischke:1995ir,Rischke:1995mt}. For the results presented here 
an equation of state for a hadron-resonance gas without any phase transition is used \cite{Zschiesche:2002zr}.
The EoS includes all hadronic degrees of freedom with masses up to $2$ GeV, which is consistent with the effective degrees of freedom present in the UrQMD model. One should note that we apply a purely hadronic EoS, for energy densities where a transition to the QGP is expected (see also \cite{Petersen:2008dd} for details on the model and comparison of extracted particle yields to data). Final particle (and MEMO) multiplicities are mainly sensitive on the degrees of freedom at chemical freezeout which is reflected in the hadronic EoS. Dynamical observables such as momentum and rapidity spectra are more sensitive on the underlying dynamics. In addition, a phase transition could catalyse a strangeness destillation process further enhancing MEMO production. However, studying the effects of a phase transition on MEMO production is left subject of future investigations.     
\\ 
\begin{table}[t]
		\begin{tabular}{|c|c|c|c|}
		\hline 
		Cluster & Mass [GeV] & Quark content\\ \hline\hline
		$He^4$& 3.750 & $12 q$ \\ \hline
		$H^0$& 2.020 & $4q + 2s$ \\ \hline
		$\alpha_q$& 6.060 & $12q+6s$\\ \hline
		$\{\Xi^-,\Xi^0\}$ & 2.634 & $2q + 4s$\\ \hline
		$\{4\Lambda\}$& 4.464 & $8q + 4s$\\ \hline
		$\{2\Xi^-,2\Xi^0\}$& 5.268 & $4q + 8s$ \\ \hline
		$^{5}_{\Lambda}He$ & 4.866 & $14q + 1s$\\ \hline
		$^{6}_{\Lambda\Lambda}He$& 5.982 & $16q + 2s$\\ \hline
	  $^{7}_{\Xi^0\Lambda\Lambda}He$& 7.297 & $16q + 2s$\\ \hline
		$\{2n,2\Lambda,2\Xi^-\}$& 6.742 & $12q + 6s$ \\ \hline
		$\{2\Lambda,2\Xi^0,2\Xi^-\}$& 7.500 & $8q + 10s$ \\ \hline
		$\{d,\Xi^-,\Xi^0\}$& 4.508 & $8q + 4s$\\ \hline
		$\{2\Lambda,2\Xi^-\}$& 4.866 & $6q + 6s$ \\ \hline
		$\{2\Lambda,2\Sigma^-\}$& 4.610 & $8q + 4s$ \\ \hline
		\end{tabular}
	\caption{Properties of all considered multibaryonic states \label{table1}}
\end{table}
\\

The hydrodynamic evolution is stopped, if the energy density of all cells drops below five times the ground 
state energy density (i.e. $\sim 730 {\rm MeV/fm}^3$). This criterion corresponds 
to a T-$\mu_B$-configuration where the phase transition is expected - approximately $T=170$ MeV 
at $\mu_B=0$. The hydrodynamic fields are mapped to particle degrees of freedom via the Cooper-Frye 
equation on an isochronous hyper-surface.
\begin{equation}
\label{cooper_frye}
E \frac{dN}{d^3p}=\int_\sigma f(x,p) p^\mu d\sigma_\mu \quad \rm{with} \quad d\sigma_{\mu}=(dx^3,\vec{0})
\end{equation}
Here $f(x,p)$ are the boosted Fermi or Bose distributions corresponding to the respective particle species. Inputs for these distributions are the masses and chemical potentials of the desired particles. For our calculation we assumed the mass of a MEMO to be the sum of the masses of all hadrons it is composed of. Similarly the total chemical potential is the sum of the constituents, and is composed of baryon and strange-quark chemical potentials $\mu_B$ and $\mu_s$.\\
The particle vector information is then transferred back 
to the UrQMD model, where rescatterings and the final decays are performed using the hadronic cascade.
Using this parametrisation of the model one obtains a satisfactory description of data in a energy regime of $1-160A$ GeV.  
A more detailed description of the hybrid model including parameter tests and results for 
multiplicities and spectra can be found in \cite{Petersen:2008dd}. 

To calculate the multiplicities of MEMOS in the FAIR energy region, we employ the introduced hybrid approach
to heavy ion collisions. 
Thus, the fluctuating initial state produced in UrQMD, 
is coupled to a (3+1) dimensional hydrodynamics evolution. When the energy density drops 
below $5\epsilon_0(\sim 730 {\rm MeV/fm}^3)$ the freeze-out is performed and MEMOs and strangelets are produced according 
to the Cooper-Frye description (\ref{cooper_frye}). As distinctive inputs for the distribution functions, the chemical potentials $(\mu_{s},\mu_B)$ and masses of the MEMOs enter as discussed above. Final state interactions of these MEMOs are neglected for the present study. Table (\ref{table1}) gives the properties of all multibaryonic states considered in our analysis. They are the most promising and stable candidates.

\begin{figure}[t]
 \centering
\includegraphics[width=0.5\textwidth]{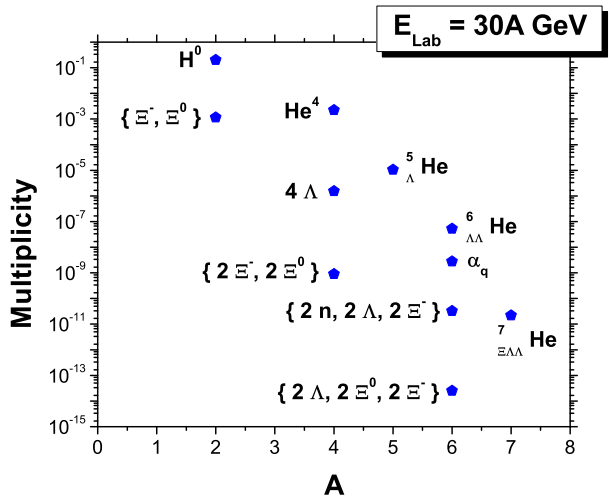}
 \caption{Multiplicities of various types of MEMOs and strangelets in central Pb+Pb reactions at $E_{\rm lab}=30 A$ GeV 
from the hybrid approach.}
 \label{fig:mul30}
\end{figure}

Fig. (\ref{fig:mul30}) provides the total multiplicities per degeneracy factor of  various types of MEMOs 
and strangelets in central Pb+Pb reactions at $E_{\rm lab}=30 A$ GeV. The yields obtained are in good comparison to
the statistical model analysis \cite{BraunMunzinger:1994iq}, which is describing strange cluster production at AGS energies.\\
One should also note that we assume particle production from a grand canonical ensemble for all beam energies.
Because local, as well as global, thermal equilibration are assumptions not necessary justified in heavy ion collisions, a microcanonical description, combined with MEMO production by coalescence, has been proposed in \cite{SchaffnerBielich:1999sy}.
Due to the restrictions of energy and momentum conservation, resulting in a phase space reduction for produced strange particles a (micro)canonical description of the system strongly decreases strange particle yields \cite{Becattini:1997rv,Cleymans:1990mn,Andronic:2005yp}.\\
On the other hand, thermal models are able to reproduce strange particle yields for beam energies above $E_{lab} \approx 8A$ GeV very well, and canonical corrections become negligible above these energies \cite{Andronic:2005yp}.\\
\\

\begin{figure}[t]
 \centering
\includegraphics[width=0.5\textwidth]{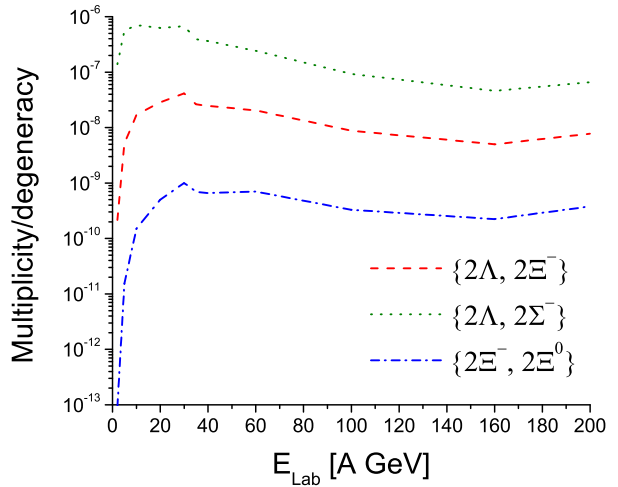}
 \caption{(color online) Excitation functions of the multiplicities of various MEMOs in central Pb+Pb reactions from the hybrid approach.}
 \label{fig:exci}
\end{figure}

Investigating strange-cluster production over a range of beam energies shows a distinct maximum in the yields of several multi strange objects. Fig. \ref{fig:exci} displays the excitation function of the multiplicities of various MEMOs in central Pb+Pb reactions from the hybrid approach. The presented MEMO candidates are expected to possess binding energies up to $E_B/A_B \approx -22 \ MeV $\cite{SchaffnerBielich:1996eh}.
One easily observes that the upper FAIR energy region ($\sim E_{\rm lab}=10-40 A$ GeV) is ideally placed
for the search of exotic multi-strange baryon clusters. At lower energies, the hyperon production cross section
is too small, while at energies above FAIR, the expansion of the source and the small baryo-chemical potential suppress the 
formation of MEMOs and strangelets. 

\begin{figure}[t]
 \centering
\includegraphics[width=0.5\textwidth]{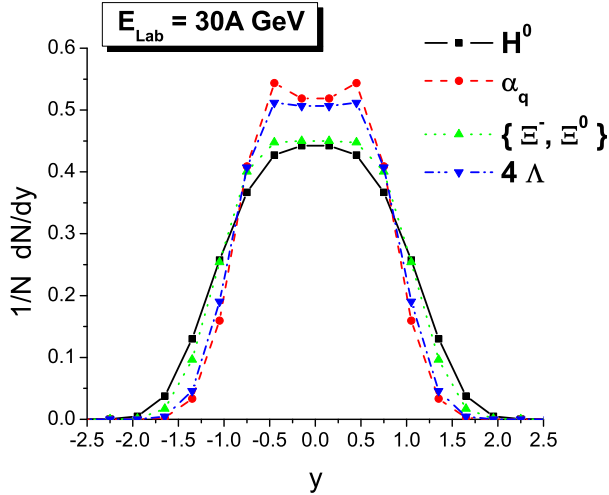}
 \caption{(color online) Normalized rapidity density of various MEMOs in central Pb+Pb reactions at 
$E_{\rm lab}=30 A$ GeV from the hybrid approach.}
 \label{fig:memo_rap30}
\end{figure}

Using the hybrid model enables us to also explore the phase space distribution of the produced particles. Fig. \ref{fig:memo_rap30} shows the
rapidity density of various MEMOs in central Pb+Pb reactions at $E_{\rm lab}=30 A$ GeV from the hybrid approach.
The production of baryon rich clusters is most pronounced in the high baryon density rapidity region. The rapidity distributions for MEMOs with a larger strangeness to baryon number fraction tend to look more gaussian like. 

\begin{figure}[t]
 \centering
\includegraphics[width=0.5\textwidth]{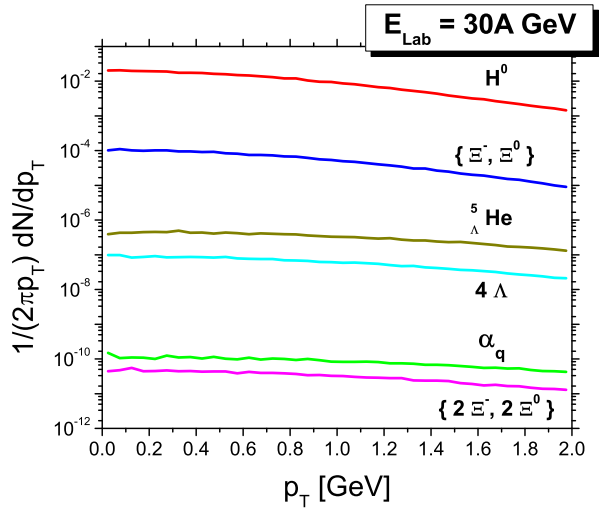}
 \caption{(color online) Transverse momentum spectra at midrapidity ($|y|<0.5$)
of various MEMOs in central Pb+Pb reactions at $E_{\rm lab}=30 A$ GeV from the hybrid approach.}
 \label{fig:memo_pt30}
\end{figure}

Figure \ref{fig:memo_pt30} depicts the transverse momentum distribution of various MEMOs at midrapidity 
in central Pb+Pb reactions at $E_{\rm lab}=30 A$ GeV from the hybrid approach. The $p_T$ spectra are rather 
broad as compared to usual hadrons. This is due to the large boost the MEMOs acquire due to their large mass and the fact, that they are produced predominantly in the hottest regions of the expanding system.

\section{Fluctuations}
For the present study so far we have assumed global as well as local strangeness conservation. These assumptions are common for models including thermal production of particles. In the following we explore if that assumption of local strangeness conservation is justified, especially in the FAIR energy regime. A relaxation of this assumption within the hybrid approach will require the explicit propagation of the strangeness density (similar to the treatment of the baryon density). A second key ingredient will be the inclusion of an equation of state that can provide $p(\epsilon,\rho_B,\rho_s)$ with a finite $\rho_s$. However, first we explore if such an extension might be necessary by applying the UrQMD model without an intermediate hydrodynamic phase.

\begin{figure}[t]
 \centering
\includegraphics[width=0.5\textwidth]{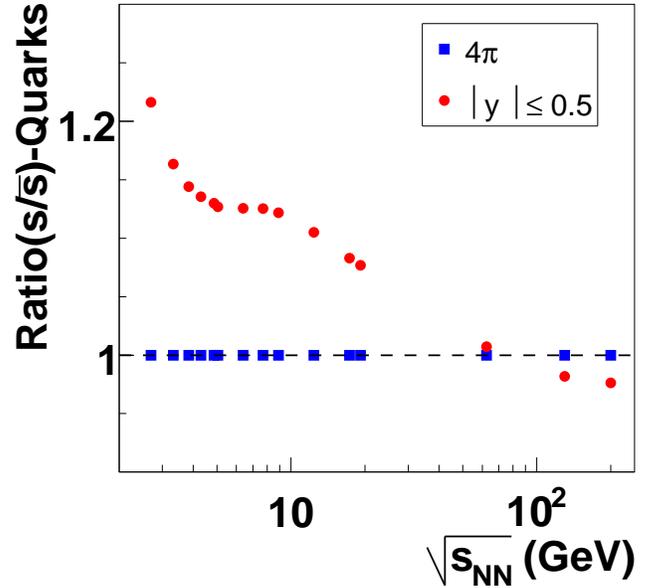}
 \caption{(color online) Energy dependence of the strange quark over anti-strange quark ($s/\overline s$) ratio for 
central Pb+Pb/Au+Au reactions.
Circles show the ratio at midrapidity, while squares show the $4\pi$ values where
the ratio is unity due to strangeness conservation.}
 \label{fig:energy_ssbar}
\end{figure}

We start with an investigation of the strangeness production and its distribution as a function of
energy. Fig. \ref{fig:energy_ssbar} depicts the energy dependence of  the strange quark over 
anti-strange quark ($s/\overline s$) ratio for central Pb+Pb/Au+Au reactions.
The red circles present the strangeness to anti-strangeness ratio at midrapidity, while the blue squares show the $4\pi$ values where the ratio is unity due to strangeness conservation. One clearly observes that strangeness is not
evenly distributed over rapidity, leading to an asymmetry between strange and anti-strange quarks on the 
level of 20\% in the relevant energy regime. A similar kind of strangeness separation process has been
predicted long ago within models coupling a hadron gas to a Quark-Gluon-Plasma 
state \cite{Greiner:1987tg,Greiner:1991us}. Within these models the energy and particle number balance in the mixed phase supports a 'distillation' process that enriches the QGP phase with strangeness and the hadronic phase with anti-strangeness. Within the present model, however, hadronic interactions are responsible for the phase space separation of strangeness and anti-strangeness since no first order phase transition is present.
Since both procedures separate strangeness in an equivalent way one can expect an even stronger strangeness separation if both effects are at work.
 
\begin{figure}[t]
 \centering
\includegraphics[width=0.5\textwidth]{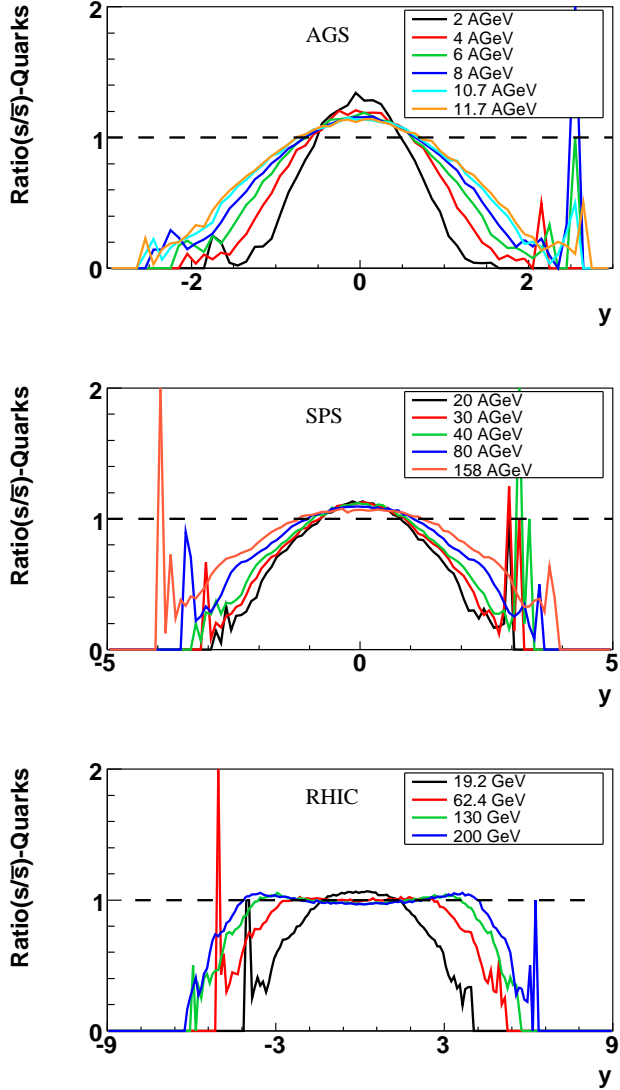}
 \caption{(color online) Rapidity dependence of the strange quark over anti-strange quark ($s/\overline s$) ratio 
for central Pb+Pb/Au+Au reactions at AGS ($E_{lab}=2-11A$ GeV), SPS ($E_{lab}=20-158A$ GeV) and 
RHIC ($\sqrt{s_{NN}}=19-200$ GeV) energies.}
 \label{fig:rapidity_ssbar}
\end{figure} 

Fig. \ref{fig:rapidity_ssbar} shows the rapidity dependence of the strange quark over 
anti-strange quark ($s/\overline s$) ratio for central Pb+Pb/Au+Au reactions at 
AGS ($E_{lab}=2-11 A$ GeV), SPS ($E_{lab}=20-158 A$ GeV) and RHIC ($\sqrt{s_{NN}}=19-200$ GeV) energies.
In the AGS and SPS energy regime, the ($s/\overline s$) ratio is strongly rapidity dependent and
has a pronounced peak above unity near midrapidity. At RHIC energies, the ($s/\overline s$) ratio turns into 
a box shape as a function of rapidity with a plateau at unity indicating that strangeness is
locally neutralized in rapidity. At the highest RHIC energies, the ($s/\overline s$) ratio even turns slightly smaller than 1.
Continuing this trend one would expect a clearly smaller than 1 ($s/\overline s$) ratio at LHC energies.
In consequence, statistical model approaches (with the constraint of strangeness conservation at mid rapidity \cite{Andronic:2005yp}) are allowed to use midrapidity particle ratios as input for their calculations only at low RHIC energies. At lower, as well as higher energies, this procedure is not justified as strangeness neutralisation does not hold for the central rapidity region.
Thermal calculations, using full phase space data as an input \cite{Becattini:2000jw,Cleymans:1999st}, and results from a thermal model including a core-corona scenario \cite{Becattini:2008ya}, generally give better descriptions of strange particle data, supporting the idea of dynamical strangeness separation, as proposed by this work.
 .

\begin{figure}[t]
 \centering
\includegraphics[width=0.5\textwidth]{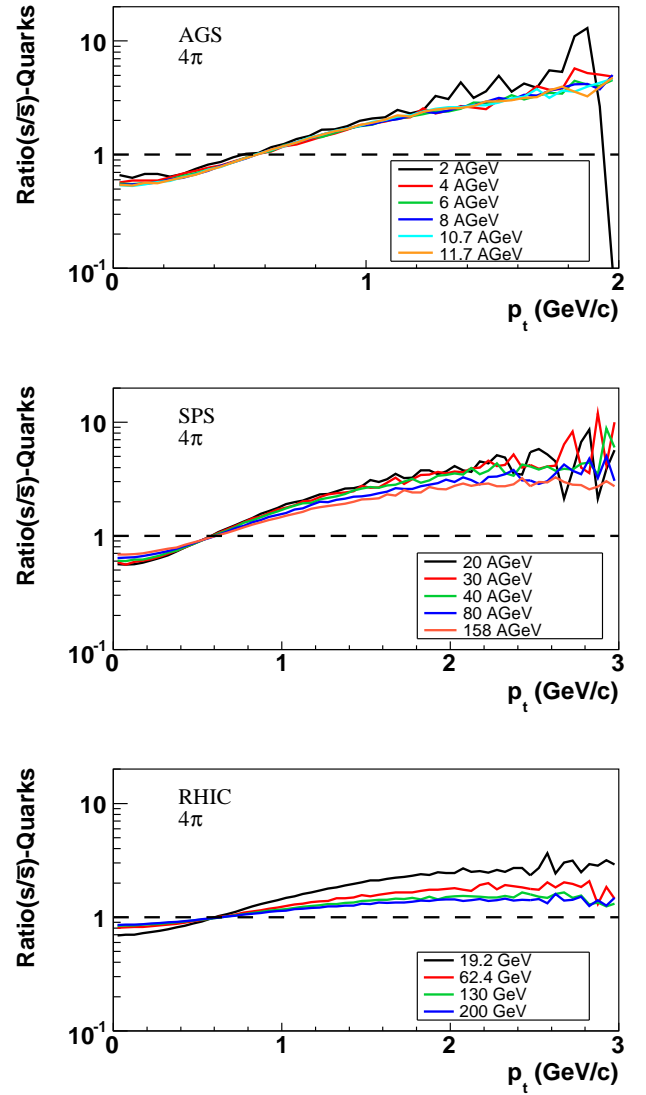}
 \caption{(color online) Transverse momentum dependence of  the strange quark over anti-strange 
quark ($s/\overline s$) ratio for central Pb+Pb/Au+Au reactions at AGS ($E_{lab}=2-11A$ GeV), 
SPS ($E_{lab}=20-158A$ GeV) and RHIC ($\sqrt{s_{NN}}=19-200$ GeV) energies.}
 \label{fig:pt_ssbar_4pi}
\end{figure}

Fig. \ref{fig:pt_ssbar_4pi} provides the transverse momentum dependence of  the strange quark 
over anti-strange quark ($s/\overline s$) ratio for central Pb+Pb/Au+Au reactions at 
AGS ($E_{lab}=2-11A$ GeV), SPS ($E_{lab}=20-158A$ GeV) and RHIC ($\sqrt{s_{NN}}=19-200$ GeV) energies.
Here one observes a strong separation of strangeness in transverse direction. With decreasing energy (increasing
baryo-chemical potential) the distribution of (anti-)strangeness becomes increasingly non-uniform in momentum space.
The low momentum region is depleted of strange quarks, while the high $p_T$ region shows a strong
enhancement of strange quarks compared to anti-strange quarks. This can be intuitively linked
to the fact that (multi)strange baryons have a larger inverse slope than the Kaons for a given transverse velocity due to their larger masses.\footnote{If one assumes strangeness conservation as well as vanishing net baryon number at midrapidity (as is expected at very high energies) then these distributions should be flat, as particles and their antiparticles are produced in equal numbers. 
If the ($s/\overline s$) ratio does deviate from unity the separation of strangeness in momentum space should still be present even at vanishing net baryon number.}  

\begin{figure}[t]
 \centering
\includegraphics[width=0.5\textwidth]{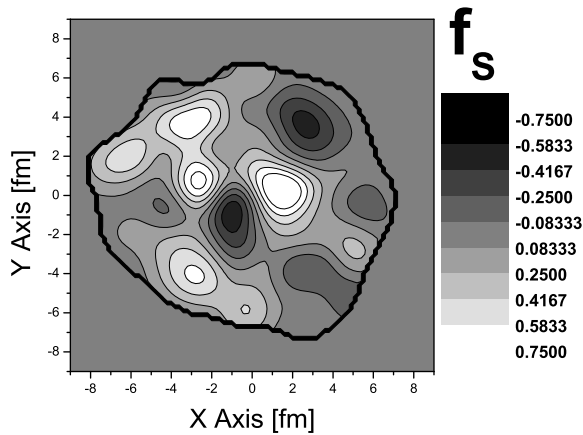}
 \caption{Fluctuations of the strangeness fraction $f_s=\rho_s/\rho_B$ 
in the central plane (x is in the impact parameter direction, y is transversal to the impact parameter
and longitudinal direction) for a single central Pb+Pb reaction at $E_{lab}=30A$ GeV. The colour 
coding indicates the local strangeness fraction, dark regions have more
anti-strange than strange quarks.}
 \label{fig:fs30}
\end{figure}

Next, we turn to the distribution and fluctuations of strangeness in coordinate space. Fig. \ref{fig:fs30} elucidates the
fluctuations of the strangeness fraction $f_s=\rho_s/\rho_B$, with $\rho_B$ being the local baryon density 
and $\rho_s$ being the local net-strangeness density, in the central plane for a single central Pb+Pb reaction 
at $E_{lab}=30A$ GeV. Here x is in the impact parameter direction and y is transversal to the 
impact parameter and longitudinal direction. The distribution of the net-strangeness and baryon densities were obtained from the UrQMD model by means similar to creating the hydro initial state in the hybrid model. All hadrons and their baryon number and strangeness content are represented by a Gaussian with a finite width of $1$ fm \cite{Steinheimer:2007iy,Petersen:2008dd}. 
The plot is shown for the time when both nuclei have passed each other. 
The colour coding indicates the local strangeness fraction, white regions have more strange 
than anti-strange quarks, while dark and black regions show more anti-strange quarks.
Locally strangeness lumps of 4~fm$^2 \times \Delta z$ appear both in positive and negative strangeness directions. 
As for the distribution in momentum space discussed above, also the coordinate space distribution is 
largely non-uniform, although these spacial fluctuations occur only on an event-by-event basis. 

We have presented results for the thermal production of MEMOs in nucleus-nucleus collisions from a combined 
micro+macro approach. Multiplicities, 
rapidity and transverse momentum spectra are predicted for Pb+Pb interaction at $E_{lab}=5A$ GeV
and $E_{lab}=30A$ GeV. The presented excitation functions for various MEMO multiplicities show a clear maximum at the upper FAIR energy regime making this facility the ideal place to study the production of these exotic forms of multistrange objects. Detector simulations have shown that the CBM experiment is well suited for the search of exotic multihypernuclear objects either by invariant mass reconstruction of strange di-baryons or observing decay systematics (The very stable double negative $\{2\Xi^0,2\Xi^-\}$ for example should have a characteristic decay in two negatively charged particles)\cite{SchaffnerBielich:1996eh}.

Compared to many previous studies on MEMO and strangelet production, based on statistical models with
global strangeness and baryon number conservation, the present approach indicates that the local strangeness density
clumps strongly in coordinate space and that strangeness is unevenly distributed in momentum space.
This mechanism does not require the production of a deconfined 
state, and profits from the non-equilibrium features present in the reaction.
These fluctuations might lead to an enhancement of MEMO (and strangelet) production compared to previous calculations.\\
Furthermore the net strangeness at midrapidity deviates from zero - not only on an event-by-event basis - indicating that the assumption of local strangeness neutralisation is only justified at the RHIC energy regime, but not at lower energies. Here it is therefore questionable if midrapidity particle ratios can be used as input for thermal particle multiplicity calculations.

\section*{Acknowledgments}

This work has been supported by GSI and Hessian initiative for excellence (LOEWE) through 
the Helmholtz International Center for FAIR (HIC for FAIR). We would like to thank 
Dr. J\"urgen Schaffner-Bielich for fruitful discussions. Computational resources were provided
by the Center for Scientific Computing (CSC). H.P. acknowledges support from the Deutsche Telekom Stiftung.


\begin{thebibliography}{99}

\bibitem{SQM98}
{International Symposium on Strangeness in Quark Matter 1998}, J. Phys. G {\bf
  25},  1  (1999).  

\bibitem{Bodmer:1971we}
  A.~R.~Bodmer,
  Phys.\ Rev.\  D {\bf 4}, 1601 (1971).

\bibitem{Jaffe:1976yi}
  R.~L.~Jaffe,
  Phys.\ Rev.\ Lett.\  {\bf 38}, 195 (1977)
  [Erratum-ibid.\  {\bf 38}, 617 (1977)].

\bibitem{Goldman:1987ma}
  J.~T.~Goldman, K.~Maltman, G.~J.~.~Stephenson, K.~E.~Schmidt and F.~Wang,
  Phys.\ Rev.\ Lett.\  {\bf 59}, 627 (1987).

\bibitem{Goldman:1998jd}
  J.~T.~Goldman, K.~Maltman, G.~J.~.~Stephenson, J.~L.~.~Ping and F.~Wang,
  Mod.\ Phys.\ Lett.\  A {\bf 13}, 59 (1998)

\bibitem{Schwesinger:1994vd}
  B.~Schwesinger, F.~G.~Scholtz and H.~B.~Geyer,
  Phys.\ Rev.\  D {\bf 51}, 1228 (1995)

\bibitem{Alt:2003vb}
  C.~Alt {\it et al.}  [NA49 Collaboration],
  Phys.\ Rev.\ Lett.\  {\bf 92}, 042003 (2004)

\bibitem{Oakes:1963zz}
  R.~J.~Oakes,
  Phys.\ Rev.\  {\bf 131}, 2239 (1963).

\bibitem{Dyson:1964}F.J. Dyson and N.H. Xuong, Phys. Rev. Lett. \textbf{13} (1964) 815.

\bibitem{Libby:1969tw}
  L.~M.~Libby,
  Phys.\ Lett.\  B {\bf 29}, 345 (1969).

\bibitem{Graffi:1969is}
  S.~Graffi, V.~Grecchi and G.~Turchetti,
  Lett.\ Nuovo Cim.\  {\bf 2S1}, 311 (1969)
  [Lett.\ Nuovo Cim.\  {\bf 2}, 311 (1969)].

\bibitem{Aerts:1977rw}
  A.~T.~M.~Aerts, P.~J.~G.~Mulders and J.~J.~de Swart,
  Phys.\ Rev.\  D {\bf 17}, 260 (1978).

\bibitem{Wong:1978tz}
  C.~W.~Wong and K.~F.~Liu,
  Phys.\ Rev.\ Lett.\  {\bf 41}, 82 (1978).

\bibitem{Aerts:1984ht}
  A.~T.~M.~Aerts and C.~B.~Dover,
  Phys.\ Lett.\  B {\bf 146}, 95 (1984).

\bibitem{Kalashnikova:1987rma}
  Yu.~S.~Kalashnikova, I.~M.~Narodetsky and Yu.~A.~Simonov,
  Sov.\ J.\ Nucl.\ Phys.\  {\bf 46}, 689 (1987)
  [Yad.\ Fiz.\  {\bf 46}, 1181 (1987)].

\bibitem{Goldman:1989zj}
  J.~T.~Goldman, K.~Maltman, G.~J.~.~Stephenson, K.~E.~Schmidt and F.~Wang,
  Phys.\ Rev.\  C {\bf 39}, 1889 (1989).

\bibitem{SchaffnerBielich:1999sy}
  J.~Schaffner-Bielich, R.~Mattiello and H.~Sorge,
  Phys.\ Rev.\ Lett.\  {\bf 84}, 4305 (2000)

\bibitem{SchaffnerBielich:2000nd}
  J.~Schaffner-Bielich,
  Nucl.\ Phys.\  A {\bf 691}, 416 (2001)

\bibitem{Ahn:2001sx}
  J.~K.~Ahn {\it et al.},
  Phys.\ Rev.\ Lett.\  {\bf 87}, 132504 (2001).

\bibitem{Takahashi:2001nm}
  H.~Takahashi {\it et al.},
  Phys.\ Rev.\ Lett.\  {\bf 87}, 212502 (2001).

\bibitem{Dalitz:1989kt}
  R.~H.~Dalitz, D.~H.~Davis, P.~H.~Fowler, A.~Montwill, J.~Pniewski and J.~A.~Zakrzewski,
  Proc.\ Roy.\ Soc.\ Lond.\  A {\bf 426}, 1 (1989)

\bibitem{Schaffner:1992sn}
  J.~Schaffner, H.~Stoecker and C.~Greiner,
  Phys.\ Rev.\  C {\bf 46}, 322 (1992).

\bibitem{Schaffner:1993nn}
  J.~Schaffner, C.~B.~Dover, A.~Gal, C.~Greiner and H.~Stoecker,
  Phys.\ Rev.\ Lett.\  {\bf 71}, 1328 (1993).

\bibitem{Gilson:1993zs}
  E.~P.~Gilson and R.~L.~Jaffe,
  Phys.\ Rev.\ Lett.\  {\bf 71}, 332 (1993)

\bibitem{SchaffnerBielich:1996eh}
  J.~Schaffner-Bielich, C.~Greiner, A.~Diener and H.~Stoecker,
  Phys.\ Rev.\  C {\bf 55}, 3038 (1997)

\bibitem{Stoks:1999bz}
  V.~G.~J.~Stoks and T.~A.~Rijken,
  Phys.\ Rev.\  C {\bf 59}, 3009 (1999)

\bibitem{Scherer:2008zz}
  S.~Scherer, M.~Bleicher, S.~Haussler and H.~St\"ocker,
  Int.\ J.\ Mod.\ Phys.\  E {\bf 17}, 965 (2008).

\bibitem{Aichelin:2008mi}
  J.~Aichelin and K.~Werner,
  arXiv:0810.4465 [nucl-th].

\bibitem{Donoghue:1986zd}
  J.~F.~Donoghue, E.~Golowich and B.~R.~Holstein,
  Phys.\ Rev.\  D {\bf 34}, 3434 (1986).

\bibitem{Krivoruchenko:1982xv}
  M.~I.~Krivoruchenko and M.~G.~Shchepkin,
  Sov.\ J.\ Nucl.\ Phys.\  {\bf 36}, 769 (1982)
  [Yad.\ Fiz.\  {\bf 36}, 1328 (1982)].

\bibitem{AlaviHarati:1999ds}
  A.~Alavi-Harati {\it et al.}  [KTeV Collaboration],
  Phys.\ Rev.\ Lett.\  {\bf 84}, 2593 (2000)

\bibitem{Belz:1995nq}
  J.~Belz {\it et al.}  [BNL-E888 Collaboration],
  Phys.\ Rev.\ Lett.\  {\bf 76}, 3277 (1996)
  [Phys.\ Rev.\  C {\bf 56}, 1164 (1997)]

\bibitem{Crawford:1998uq}
  H.~J.~Crawford,
  Nucl.\ Phys.\  A {\bf 639}, 417 (1998).

\bibitem{Caines:1999ia}
  H.~Caines {\it et al.}  [E896 Collaboration],
  Nucl.\ Phys.\  A {\bf 661}, 170 (1999).

\bibitem{Appelquist:1996qx}
  G.~Appelquist {\it et al.}  [NA52 (NEWMASS) Collaboration],
  Phys.\ Rev.\ Lett.\  {\bf 76}, 3907 (1996).

\bibitem{Armstrong:1997sc}
  T.~A.~Armstrong {\it et al.}  [E864 Collaboration],
  Phys.\ Rev.\ Lett.\  {\bf 79}, 3612 (1997)

\bibitem{Finch:1999ja}
  L.~E.~Finch  [E864 Collaboration],
  Nucl.\ Phys.\  A {\bf 661}, 395 (1999)

\bibitem{Stoecker:1994ud}
  H.~Stoecker {\it et al.},
  Nucl.\ Phys.\  A {\bf 566}, 15C (1994).

\bibitem{Sorge:1989dy}
  H.~Sorge, H.~Stoecker and W.~Greiner,
  Annals Phys.\  {\bf 192}, 266 (1989).

\bibitem{NilssonAlmqvist:1986rx}
  B.~Nilsson-Almqvist and E.~Stenlund,
  Comput.\ Phys.\ Commun.\  {\bf 43}, 387 (1987).

\bibitem{Andersson:1986gw}
  B.~Andersson, G.~Gustafson and B.~Nilsson-Almqvist,
  Nucl.\ Phys.\  B {\bf 281}, 289 (1987).

\bibitem{Sjostrand:1993yb}
  T.~Sjostrand,
  Comput.\ Phys.\ Commun.\  {\bf 82}, 74 (1994).

\bibitem{Bass:1998ca}
  S.~A.~Bass {\it et al.},
  Prog.\ Part.\ Nucl.\ Phys.\  {\bf 41}, 255 (1998)
  [Prog.\ Part.\ Nucl.\ Phys.\  {\bf 41}, 225 (1998)]

\bibitem{Bleicher:1999xi}
  M.~Bleicher {\it et al.},
  J.\ Phys.\ G {\bf 25}, 1859 (1999)

\bibitem{Steinheimer:2007iy}
  J.~Steinheimer, M.~Bleicher, H.~Petersen, S.~Schramm, H.~St\"ocker and D.~Zschiesche,
  Phys.\ Rev.\  C {\bf 77}, 034901 (2008)

\bibitem{Rischke:1995ir}
  D.~H.~Rischke, S.~Bernard and J.~A.~Maruhn,
  Nucl.\ Phys.\  A {\bf 595}, 346 (1995)

\bibitem{Rischke:1995mt}
  D.~H.~Rischke, Y.~Pursun and J.~A.~Maruhn,
  Nucl.\ Phys.\  A {\bf 595}, 383 (1995)
  [Erratum-ibid.\  A {\bf 596}, 717 (1996)]

\bibitem{Zschiesche:2002zr}
  D.~Zschiesche, S.~Schramm, J.~Schaffner-Bielich, H.~Stoecker and W.~Greiner,
  Phys.\ Lett.\  B {\bf 547}, 7 (2002)

\bibitem{Petersen:2008dd}
  H.~Petersen, J.~Steinheimer, G.~Burau, M.~Bleicher and H.~St\"ocker,
  Phys.\ Rev.\  C {\bf 78}, 044901 (2008)

\bibitem{BraunMunzinger:1994iq}
  P.~Braun-Munzinger and J.~Stachel,
  J.\ Phys.\ G {\bf 21}, L17 (1995)

\bibitem{Becattini:1997rv}
  F.~Becattini and U.~W.~Heinz,
  Z.\ Phys.\  C {\bf 76}, 269 (1997)
  [Erratum-ibid.\  C {\bf 76}, 578 (1997)]

\bibitem{Cleymans:1990mn}
  J.~Cleymans, K.~Redlich and E.~Suhonen,
  Z.\ Phys.\  C {\bf 51}, 137 (1991).


\bibitem{Andronic:2005yp}
  A.~Andronic, P.~Braun-Munzinger and J.~Stachel,
  Nucl.\ Phys.\  A {\bf 772}, 167 (2006)

\bibitem{Greiner:1987tg}
  C.~Greiner, P.~Koch and H.~Stoecker,
  Phys.\ Rev.\ Lett.\  {\bf 58}, 1825 (1987).

\bibitem{Greiner:1991us}
  C.~Greiner and H.~Stoecker,
  Phys.\ Rev.\  D {\bf 44}, 3517 (1991).

\bibitem{Becattini:2000jw}
  F.~Becattini, J.~Cleymans, A.~Keranen, E.~Suhonen and K.~Redlich,
  Phys.\ Rev.\  C {\bf 64}, 024901 (2001)

\bibitem{Cleymans:1999st}
  J.~Cleymans and K.~Redlich,
  Phys.\ Rev.\  C {\bf 60}, 054908 (1999)

\bibitem{Becattini:2008ya}
  F.~Becattini and J.~Manninen,
  Phys.\ Lett.\  B {\bf 673}, 19 (2009)

\end{thebibliography}
\end{document}